\documentclass[aps,prl,preprint,showpacs]{revtex4}

\usepackage{graphicx}
\usepackage{dcolumn}

\begin{document}

\title{Nanowires and Suspended Atom Chains from Metal alloys}

\author{J. Bettini$^{1}$}
\author{F. Sato$^{2}$}
\author{P.Z. Coura$^{3}$}
\author{S.O. Dantas$^{3}$}
\author{D.S. Galv\~{a}o$^{2}$}
\author{D. Ugarte$^{1,2}$}
\email{ugarte@lnls.br}

\affiliation{$^{1}$Laborat\'{o}rio Nacional de Luz S\'{\i}ncrotron, CP 6192,
13084-971 Campinas, SP, Brazil}
\affiliation{$^{2}$Instituto de F\'{\i}sica Gleb
Wataghin, Universidade Estadual de Campinas, CP 6165, 13083-970
Campinas, SP, Brazil}
\affiliation{$^{3}$Departamento de F\'{\i}sica,
ICE, Universidade Federal de Juiz de Fora, 36036-330 Juiz de Fora,
MG, Brazil}
\date{\today}

\begin{abstract}

We present a study of the elongation and rupture of gold-silver alloy nanowires. Atomistic details of the evolution were derived from time-resolved atomic resolution transmission electron microscopy and molecular dynamics simulations. The results show the occurrence of gold enrichment at the nanojunction region, leading to a gold-like structural behavior even for alloys with minor gold content. Our observations have also revealed the formation of mixed (Au and Ag) linear atomic chains.

\end{abstract}


\maketitle

For materials in the nanometric range, size reduction and high surface-volume ratio induce novel chemical and physical properties quite different from the corresponding bulk materials; as a consequence huge effort is being addressed to study and exploit nanosystems. However, most studies have provided understanding on pure nanosized substances or phases and very little is known on nanometric alloy materials. Basic aspects, such as, preferred atomic arrangement, phase stability, surface induced chemical composition gradients, etc. represent still open questions with reduced quantitative or even qualitative description \cite{mori}. For example, in the field of nanoelectronics, a large amount of theoretical and experimental work is at present addressed to study atomic-size pure metal nanowires (NWs) generated by mechanical stretching (see recent review by Agra\"{i}t \textit{et al}. \cite{revagra}), however rather limited information has been obtained on the electrical and mechanical behavior of metal alloy nanowires \cite{sakai, bakker, jan, fujii}. These issues will be very important to understand and produce the electric contacts and interconnects of nanodevices.

In this work, we have theoretically and experimentally analyzed the atomistic aspects of the elongation and rupture of Au-Ag alloys NWs. We have observed an evolution of the chemical composition of the NW region during elongation, leading to predominance of gold-like NW behavior in the alloy. Also, our results have revealed the formation of mixed, Au and Ag, suspended linear atom chains (LACs).

Metal NWs were produced \textit{in situ} in the HRTEM (JEM-3010 URP 300kV, 0.17 nm point resolution) using the methodology proposed by Kondo and Takayanagi \cite{kontak}. Initially holes are opened at several points in a self-supported metal film by focusing the microscope electron beam ($\sim$120 A/cm$^{2}$). When two holes become very close, nanometric constrictions (bridges) are formed between them. Then, the microscope beam current density is reduced to $\sim$10-30 A/cm$^{2}$ for image acquisition \cite{vrbook}, and the nanowires evolve spontaneously, elongate and finally break, sometimes with the appearance of atomic suspended chains. These processes are registered using a high sensitive TV camera (Gatan 622SC, 30 frames/s) and a standard video recorder.

The Au-Ag alloy thin films films (10-30 nm in thickness) were prepared by thermal co-evaporation of both metals in a standard vacuum evaporator (10$^{-7}$ mbar). A quartz crystal monitor was used to set the evaporation rate of each metal source and, subsequently, to measure the equivalent thickness of the alloy film. This process is quite simple and reproducible, allowing the preparation of alloy films over a large range of compositions. During the intense electron beam irradiation used to produce the NWs inside the HRTEM, the alloy nominal composition change to a minor Ag content due to Au/Ag different thermal diffusion rates. Thus, it must be noticed that the alloy compositions reported here were derived from energy dispersive x-ray analysis after electron irradiation and not to the as-deposited ones.

The elongation/rupture of pure Au or Ag NWs have been already analyzed in details by Rodrigues \textit{et al}. \cite{vrprl, regoprb,vrag} using real time HRTEM. Au and Ag are both metals with fcc structure (face centered cubic) and with similar lattice parameters, then it is not surprising that their NW behaviors share some structural features. For example, wires adopt merely three configurations, where the elongation direction follows the [100], [110] and [111] crystal axes. Notwithstanding, clear differences between Au and Ag NWs stretching behavior were observed, which are related to subtle changes of the surface energy of different crystal facets \cite{marks,vrbook}. Briefly, two clear differences can be distinguished; firstly,  NWs along [110] axis (hereafter noted [110] NWs) show a rod-like morphology with aspect-ratio in the 3-6 range for Au and a  higher aspect ratio (5-12) for Ag. Also, Au [110] NWs break abruptly (as being brittle) when they are 3-4 atom layer thick, while Ag [110] wires evolve to form LACs. Secondly, [111] NWs show by-piramidal shape for both metals, but Au ones evolve to form atom chains, while Ag ones break abruptly without LAC formation \cite{vrprl,vrag,jeffappa}.
Considering the quite differentiated behavior of pure Au and Ag NWs for the [110] and [111] directions, they are a natural choice to analyze how alloy composition affect NW formation and evolution.

In general terms, the analysis of metal alloy NWs revealed that unlike pure metals structural defects (mainly twins and stacking faults) are sometimes present at the apexes and very close to the narrowest wire constrictions or even in the NW themselves. This fact is expected because the presence of different atoms (one species acting as impurity) tend to produce what is called solid solution strengthening in metal alloys \cite{callister}.

Fig. 1 shows typical results for rod-like [110] NWs generated from different alloy compositions. We can observe a similar pillar-like morphology but with a rather low aspect ratio. This kind of morphology is typical of pure Au NWs and it is observed even for a rather low Au content (20\%); also, the typical brittle behavior of Au [110] wires is observed in spite of a high Ag content. This can be better visualized in the complementary material (video 01, Au$_{60}$Ag$_{40}$ NW, associated with Fig. 1b) \cite{video}. Fig. 2 displays some examples of LACs formed by the elongation of alloy nanowires. For example, Fig. 2b shows a LAC formed from Au$_{60}$Ag$_{40}$ alloy junction elongated along [111] and, it must be stressed that pure Ag [111] NWs break abruptly without LAC formation.

Although, we could expect alloy NWs to exhibit a gradual evolution of mixed Au and Ag behavior, this is only observed for rod-like wires with low Au content; a high Ag atomic percentage ($\sim$80\%) is necessary to reveal typical silver features.  For example, the rod-like Au$_{20}$Ag$_{80}$  wire (Fig. 1d) shows the typical low aspect-ratio of Au NWs, but the rupture is not brittle-like; in fact, this [110] NW evolves to form a suspended atomic chain (shown in Fig. 2d), as observed for pure Ag [110] NWs (see complementary material \cite{video}, video 02). The dominant gold structural behavior for elongation and rupture of the NWs is consistent with conclusion derived from quantum measurements of the same system \cite{ sakai, bakker}.

The images in Fig. 2 represent the first experimental evidences that LAC formation is possible from metallic alloys. One intriguing question is about the nature of these atom chains: Are they pure or composed by different atomic species? In this sense, we have observed that experimental images of metal alloy chains frequently show LACs formed by three atoms with a darker atom at one end (see Fig. 2d).

We must consider that HRTEM images provide a bi-dimensional view (i.e. a projection) of the tri-dimensional atomic arrangement. For such tiny structures as the wires studied in this work and provided that the images are taken at Scherzer defocus \cite{carter}, the expected contrast of the atomic columns should be proportional, in a first approximation, to the projected atomic potential (or to the number of atoms along the observational direction). At Scherzer defocus atomic columns are represented by dark spots, thus darker spots indicate thicker atom columns. A careful analysis of this kind of contrast changes may allow a more quantitative interpretation of NW image contrast. For example, Bettini \textit{et al}. \cite{jeffappa} have been able to derive detailed information on the tri-dimensional atomic arrangement evolution of rod-like Ag NWs. Here, we have used the same approach to analyze the chemical nature (Au or Ag) of the suspended atoms.

Fig. 3b shows the experimental LAC HRTEM intensity profile from the Au$_{20}$Ag$_{80}$ alloy (Fig. 2d), the darker atom (or deeper valley) can be easily observed at LAC right end. We have compared the experimental profiles with simulated image intensities \cite{JEMS} for different test case LACs formed by: (a) pure silver, three atom chain (Ag-Ag-Ag); (b) a mixed chain including a Au atom at the right (Ag-Ag-Au) (simulated HRTEM image in Fig. 3a) and, finally, (c) pure silver, but with two Ag atoms along the observation direction at the right end (Ag-Ag-2Ag; this may also generate a darker spot in the image). The second configuration reproduces the experimental contrast profile while the other two configurations differ significantly. We have analyzed several different experiments where a darker suspended atom is observed and, the extracted intensity profiles always show the same intensity pattern and contrast changes in excellent agreement with the image simulation. These results strongly suggest that the LACs can be formed as a mixture of Au and Ag atoms.

In order to obtain more insights on the structural evolution and dynamical aspects of the metal alloy NWs, we have also carried out molecular dynamics simulations where we can track the position of each atom during the elongation and rupture. We have used tight-binding molecular dynamics methodology using second-moment approximation (TB-SMA) \cite{CR,TAB}. This methodology has recently proved to be very effective for the study of Au and Cu NWs \cite{coura2004,sato2005,JuanCu}. Details of the methodology were published elsewhere \cite{sato2005}.

We have carried out simulations in the temperature range of 300-400 K, number of atoms varying between 400-600 depending on the crystallographic orientation and, the pulling velocity varying between 0.1-1.5 m/s. We have analyzed several different alloy concentrations. In order to evaluate the variation of Ag and Au as function of time evolution, we divided the NW into three regions (3/3, 2/3, 1/3 measured from the NW center, excluding the buffer layers), the central one corresponding to the region where the LACs are formed. The results show that there is a clear tendency of gold nominal enrichment in the final stages of the NW stretching. In Fig. 4 we show a typical example of this phenomenon. We must emphasize that gold concentration in the neck region tend to increase even at low gold concentration; this enrichment may be due to gold surface segregation.

The experimental data of Figs. 1 and 2 showed that the alloy structures tend to exhibit the behavior of pure Au NWs, even for small Au content. The results from MD simulations show that this can be explained due to the gold enrichment on the final stages of NW elongation. Another important point from the MD simulations was the formation of suspended chains with mixed atoms (see Fig. 5 and complementary material, video 03 \cite{video}). This is consistent with the experimental results from the Fig. 3 and an independent validation on the possibility existence of mixed LACs.

In summary, we have presented the experimental and theoretical evidences that atomic-size Au-Ag alloy nanowires generated by mechanical stretching show a gold enrichment of the nanojunction region. This leads to a dominant gold like behavior for the alloy NWs elongation and rupture, even for alloys with minor gold content. Finally, our results have also revealed the formation of mixed (Au and Ag) linear atomic chains.
\begin{acknowledgments}
This work was supported by LNLS, CNPq, FAPESP, IMMP/MCT, IN/MCT and CAPES. The authors acknowledge the invaluable help of the LNLS staff, in particular P.C. Silva for sample preparation.
\end{acknowledgments}

\pagebreak 

\begin{figure}[httb]
\includegraphics[width = 8 cm]{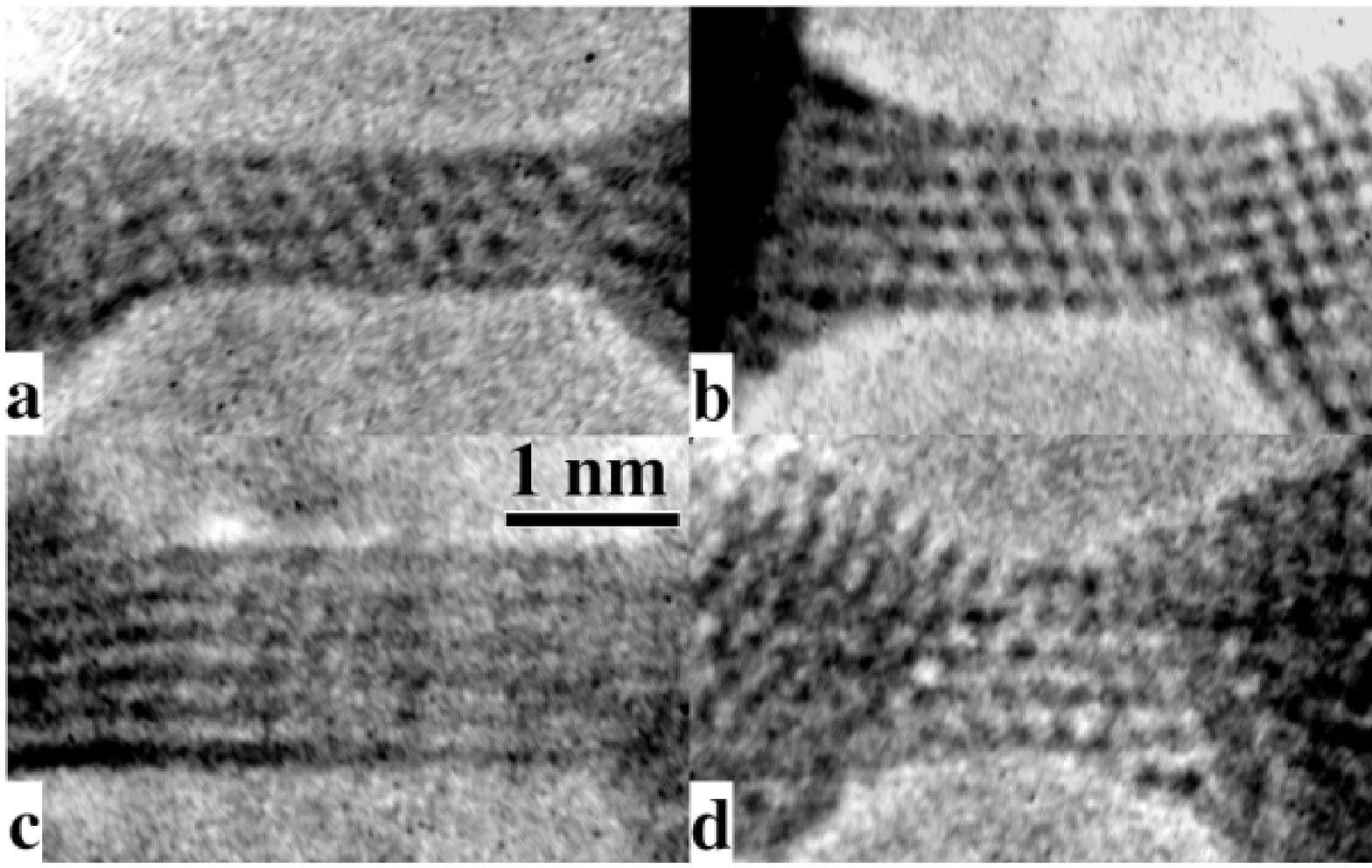}
\caption{}
\end{figure}
a-d) Atomic resolved HRTEM snapshots of rod-like [110] NWs generated from different alloy compositions Au$_{1-x}$Ag$_{x}$, for x=0.2; 0.4; 0.6 and 0.8, respectively. Atomic positions appear dark. See text for discussion.

\pagebreak 

\begin{figure}[httb]
\includegraphics[width = 8 cm]{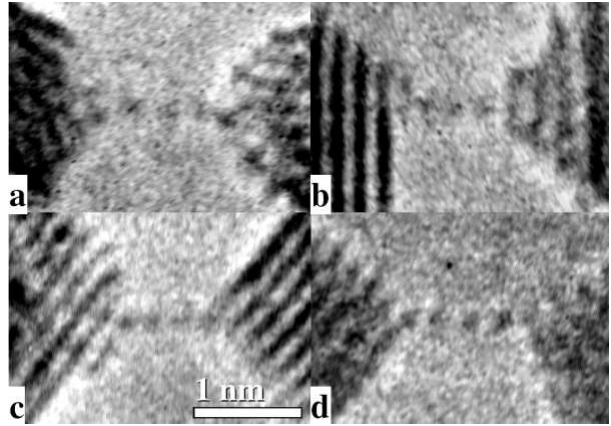}
\caption{ a-d) Atomic resolved HRTEM snapshots of suspended atom chains generated from different alloy compositions Au$_{1-x}$Ag$_{x}$, for x=0.2; 0.4; 0.6 and 0.8, respectively. Atomic positions appear dark. See text for explanations. }
\end{figure}

\pagebreak 

\begin{figure}[httb]
\includegraphics[width = 8 cm]{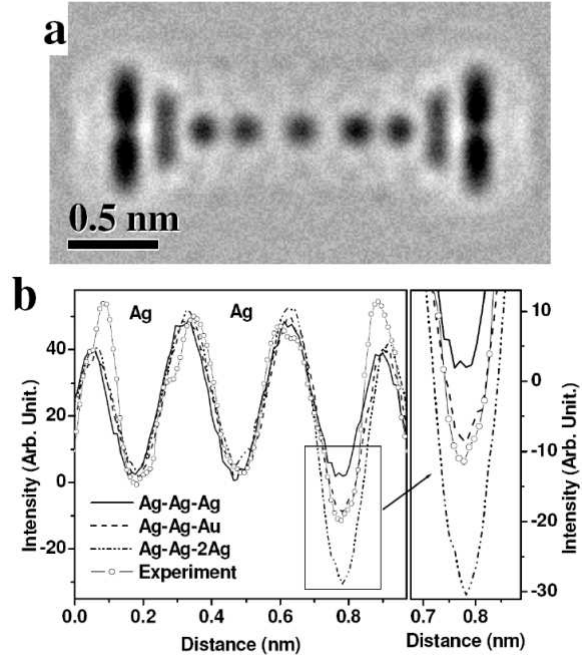}
\caption{ a) HRTEM simulated image of a mixed atom chain formed by three suspended atoms hanging between to Ag tips, the LAC is formed two Ag atoms and a Au one at the right end. b) Comparison between experimental (Au$_{20}$Ag$_{80}$) and simulated intensity profiles across a metal alloy LAC formed by three hanging atoms (see text for discussions). Atomic positions appear as intensity minima and are indicated by arrows in the profile. }
\end{figure}

\pagebreak 

\begin{figure}[httb]
\includegraphics[width = 8 cm]{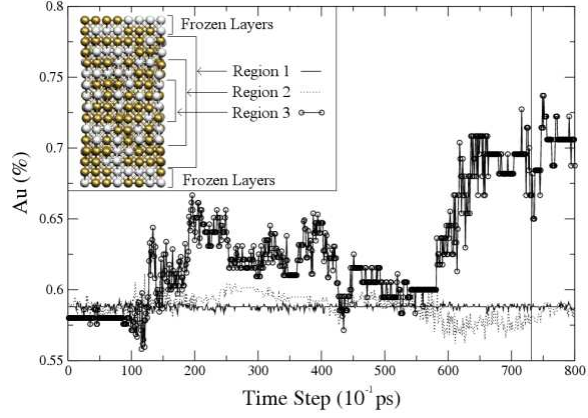}
\caption{Nominal percentage of Au atoms as a function of time evolution for the three different regions of the alloy NW (Au$_{60}$Ag$_{40}$ see text for explanationis). Inset: initial configuration of the molecular dynamics simulation. Yellow (gray) indicate gold (silver) atoms. The different regions used for measuring the concentrations are also indicated. It is clear the increase in the gold percentage in the final elongation steps. The vertical line at 729 time step indicates the moment of NW rupture. See text for discussions.}
\end{figure}

\pagebreak 

\begin{figure}[httb]
\includegraphics[width = 6 cm]{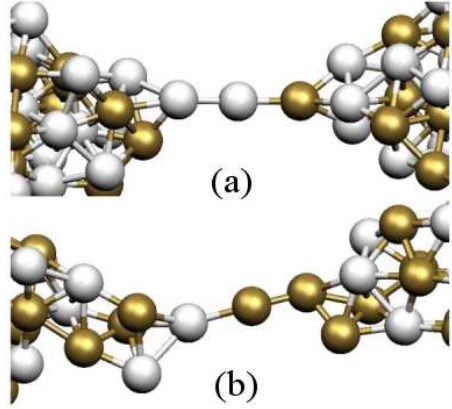}
\caption{Snapshots from the molecular dynamcs simulations showing the formation of mixed LACs from alloys: (a) Au$_{40}$Ag$_{60}$ and (b) Au$_{60}$Ag$_{40}$. Yellow (gray) indicate gold (silver) atoms.}
\end{figure}

\end{document}